\documentclass[preprint, superscriptaddress,amsmath, nofootinbib]{revtex4-1}
\usepackage{graphicx}
\usepackage{latexsym}
\usepackage{bm}
\usepackage{color}
\usepackage{url}
\usepackage[colorlinks,citecolor=blue,urlcolor=blue,linkcolor=blue]{hyperref}
\usepackage{diagbox}

\usepackage{soul}
\usepackage{ulem}
\soulregister\cite7
\soulregister\ref7
\usepackage{cleveref}

\begin{document}

\title{Probing triple Higgs coupling with machine learning at the LHC}

	\author{Murat Abdughani}
	\email{mulati@pmo.ac.cn}
	\affiliation{Key Laboratory of Dark Matter and Space Astronomy, Purple Mountain Observatory, Chinese Academy of Sciences, Nanjing 210023, China}

	\author{Daohan Wang}
	\email{wangdaohan@mail.itp.ac.cn}
	\affiliation{CAS Key Laboratory of Theoretical Physics, Institute of Theoretical Physics, Chinese Academy of Sciences, Beijing 100190, China}
	\affiliation{School of Physics, University of Chinese Academy of Sciences, Beijing 100049, China}
	
	\author{Lei Wu}
	\email{leiwu@njnu.edu.cn}
	\affiliation{Department of Physics and Institute of Theoretical Physics, Nanjing Normal University, Nanjing, 210023, China}
	
	\author{Jin Min Yang}
	\email{jmyang@itp.ac.cn}
	\affiliation{CAS Key Laboratory of Theoretical Physics, Institute of Theoretical Physics, Chinese Academy of Sciences, Beijing 100190, China}
	\affiliation{School of Physics, University of Chinese Academy of Sciences, Beijing 100049, China}

	\author{Jun Zhao}
	\email{zhaojun@mail.itp.ac.cn}
	\affiliation{CAS Key Laboratory of Theoretical Physics, Institute of Theoretical Physics, Chinese Academy of Sciences, Beijing 100190, China}
	\affiliation{School of Physics, University of Chinese Academy of Sciences, Beijing 100049, China}

\begin{abstract}
Measuring the triple Higgs coupling is a crucial task in the LHC and future collider experiments. We apply the Message Passing Neural Network (MPNN) to the study of non-resonant Higgs pair production process $pp \to hh$ in the final state with $2b + 2\ell + E_{\rm T}^{\rm miss}$ at the LHC. Although the MPNN can improve the signal significance, it is still challenging to observe such a process at the LHC. We find that a $2\sigma$ upper bound (including a 10\% systematic uncertainty) on the production cross section of the Higgs pair is 3.7 times the predicted SM cross section at the LHC with the luminosity of 3000 fb$^{-1}$, which will limit the triple Higgs coupling to the range of $[-3,11.5]$.
\end{abstract}

\maketitle

\tableofcontents

\newpage
\section{Introduction}
	
The discovery of a  125 GeV Higgs boson~\cite{Aad:2012tfa,Chatrchyan:2012xdj} is a great leap in the quest to the origin of mass. The precision measurement of the Higgs couplings is one of the primary goals of the LHC experiment, which will further reveal the electroweak symmetry breaking mechanism and shed lights on the new physics beyond the Standard Model (SM). Although the current measurements of the Higgs couplings with fermions and gauge bosons are compatible with that predicted by the SM, testing the triple and quartic Higgs self-interactions is rather challenging at the LHC~(for recent reviews, see e.g.,~\cite{DiMicco:2019ngk,Dawson:2018dcd,Spira:2016ztx,Torassa:2018eng,Maas:2017wzi,Rappoccio:2018qxp, Baglio:2012np,Dolan:2012rv,Englert:2014uua,Papaefstathiou:2015paa,Chen:2015gva,Fuks:2015hna,Kilian:2017nio,Agrawal:2017cbs,Fuks:2017zkg}).

In the Brout-Englert-Higgs mechanism of electroweak symmetry breaking~\cite{Englert:1964et,Higgs:1964pj,Higgs:1966ev,Kibble:1967sv,Guralnik:1964eu}, the Higgs boson is a massive scalar with  self-interactions. The Higgs self-couplings are determined by the structure of the scalar potential,
\begin{equation}\label{eq:higgs}
	V = \frac{m_h^2}{2} h^2 + \lambda_3 v h^3 + \frac{1}{4} \lambda_4 h^4 ~,
\end{equation}
where $m_h$ is the mass of the SM Higgs boson and $v$ is the vacuum expectation value of the SM Higgs field. The $\lambda_3$ and $\lambda_4$ are the Higgs self-couplings, and the corresponding SM values are
\begin{equation}
	\lambda_3^{\rm SM} = \lambda_4^{\rm SM} = \frac{m_h^2}{2 v^2} .
\end{equation}

The values of $\lambda_3$ and $\lambda_4$ are measured via the double and triple Higgs production processes, respectively. In many extensions of the SM, these couplings can be altered by Higgs mixing effects or higher order corrections induced by new particles, such as Two Higgs Doublet Model~\cite{Kanemura:2002vm,Kanemura:2004mg,1797260} and (Next-to-)Minimal Supersymmetric Standard Model~\cite{Dobado:2002jz,Brucherseifer:2013qva,Nhung:2013lpa,Wu:2015nba}. Since the Higgs self-coupling plays an important role in vacuum stability~\cite{Degrassi:2012ry} and electroweak baryogenesis~\cite{Kobakhidze:2015xlz,Huang:2015izx}, measuring the Higgs self-coupling will provide a crucial clue to new physics~\cite{Efrati:2014uta}.

\begin{figure}[ht]
  \centering
  \includegraphics[width=15cm]{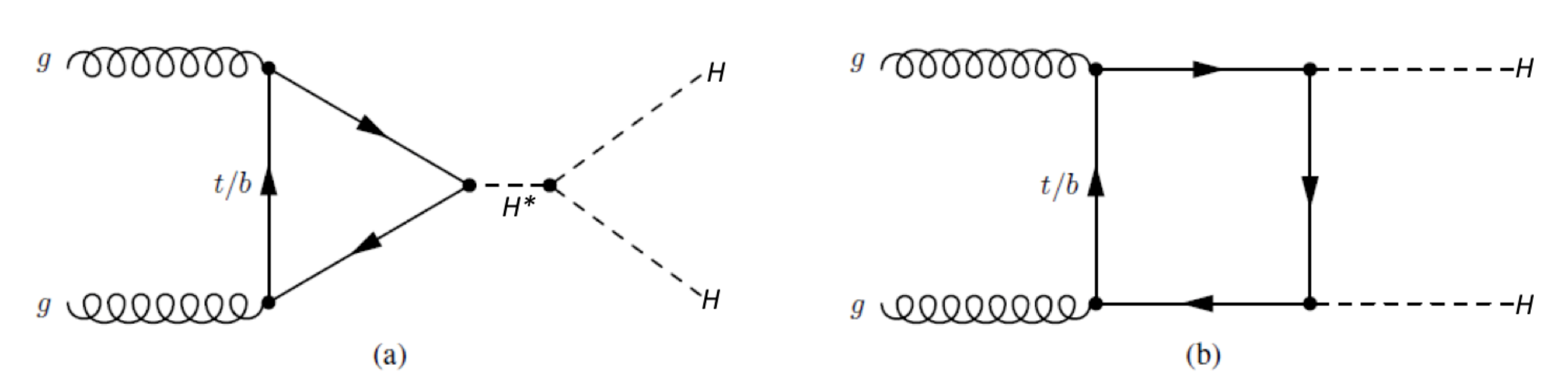}
 \caption{The representative Feynman diagrams for the Higgs boson pair production in the SM: (a) through triple Higgs
self-coupling; (b) through Higgs-fermion Yukawa interaction.}
\label{fig:higgs}
\end{figure}

The triple Higgs coupling can be indirectly probed by using the loop effects in some observables, for example, the single Higgs production~\cite{McCullough:2013rea,Gorbahn:2016uoy,Maltoni:2017ims}, and the electroweak precision observables~\cite{Kribs:2017znd}. With 80 fb$^{-1}$ of LHC Run-2 data, the triple Higgs coupling has been constrained in the range $-3.2<\lambda_3/\lambda_3^{SM}<11.9$ at 95\% C.L.~\cite{ATL-PHYS-PUB-2019-009}. On the other hand, the Higgs pair production provides a direct way to measure the triple Higgs coupling at the LHC. Such a production is dominated by the gluon-gluon fusion process, which has two main contributions: one is from the triangle diagram induced by the triple Higgs coupling, and the other is from the box diagram mediated by the top quark, as shown in Fig.~\ref{fig:higgs}. It should be noted that these two amplitudes interfere destructively, and thus results in a small cross section of $38.65$ fb for the production process $gg \to hh$ at 14 TeV LHC, which is computed at next-to-next-to-next-to-leading order (N$^{3}$LO) and including finite top quark mass effects~\cite{Chen:2019fhs}. The new physics effects that can significantly modify the Higgs pair production have been intensively studied at the LHC (see, for examples~\cite{Dolan:2012ac,Abe:2012fb,Han:2013sga,Wang:2013jwa,Hespel:2014sla,Dawson:2015oha,Kobakhidze:2016mfx,Lu:2015qqa,Ren:2017jbg,Borowka:2018pxx,Wu:2019hso,Alves:2018oct} and references therein).

In Refs.~\cite{Dolan:2012rv,Baglio:2012np,Barr:2013tda,Li:2013flc,Cao:2015oaa,Li:2015yia,Cao:2015oxx,Cao:2016zob,He:2015spf,Huang:2017jws,Lu:2015jza,Chang:2018uwu,Papaefstathiou:2012qe,Kim:2018cxf,Buchalla:2018yce,Kim:2019wns,Li:2019uyy,Sirunyan:2017guj,Aaboud:2018zhh}, the potential of measuring the Higgs pair production has been investigated in various decay modes:
$b \bar b b \bar b$, $b \bar b \tau^+ \tau^-$, $b \bar b W W^*$, $\gamma \gamma b \bar b$,
$\gamma \gamma W W^*$ and $W W^* W W^*$. Among these channels, the process of $hh \to 4b$ has the largest branching ratio, while the process of $hh \to b\bar{b}\gamma\gamma$ has a more promising sensitivity because of the low backgrounds. Using the combination of the above six analyses, the ratio $-5<\lambda_3/\lambda_3^{SM}<12$ is constrained at 95\% C.L. at 13 TeV LHC with the luminosity of 36.1 fb$^{-1}$~\cite{Aad:2019uzh}.  The sensitivity will be greatly improved at the HL-LHC~\cite{Cepeda:2019klc} and future hadron colliders~\cite{Contino:2016spe}.

In this paper, we focus on the Higgs pair production at the HL-LHC with 3ab$^{-1}$ luminosity, where one Higgs decays to $b\bar b$ and the other to $WW^*$. The decay branching ratio of $hh \to WW^*$ is the second largest after $hh \to b\bar b$, so the $b\bar b WW^*$ final state can thus act as an important channel to enhance the combining result if the signal can be well separated from the dominant $t\bar t$ background. Earlier conventional cut-flow analyses~\cite{Huang:2017jws,Papaefstathiou:2012qe,Aaboud:2018zhh} and machine learning methods~\cite{Adhikary:2017jtu,Sirunyan:2017guj,CMS:2015nat,CMS:2017cwx} applied to this channel get no more than 1 $\sigma$ significance, while recent work~\cite{Kim:2019wns} have combined the Deep Neural Network (DNN) and Convolutional Neural Network (CNN) methods to reach the significance of 1 $\sigma$. Given the importance of the Higgs pair production, in this work apply the machine learning method Message Passing Neural Network (MPNN)~\cite{DBLP:journals/corr/GilmerSRVD17}  to explore the potential of observing such di-Higgs events through the channel $pp \to hh \to b \bar b W W^*$.

In addition to the conventional kinematic cut-flow analyses, the machine learning methods have been proposed to accelerate the discovery of new physics~\cite{Roe:2004na, Baldi:2014kfa, Baldi:2014pta, Bridges:2010de, Buckley:2011kc, Bornhauser:2013aya, Caron:2016hib, Bertone:2016mdy, Abdughani:2019wuv, Bhat:2010zz, Ren:2017ymm, Lim:2018toa, Amacker:2020bmn, Hajer:2018kqm, Andreassen:2020nkr, Bishara:2019iwh, Li:2020vav, Mikuni:2020wpr, Mullin:2019mmh, Jin:2019cbv, Moreno:2019neq, Moreno:2019bmu, Jung:2019iii, Bhattacherjee:2019fpt, Qasim:2019otl, Martinez:2018fwc, Komiske:2018cqr, Farina:2018fyg, Bothmann:2018trh, Lin:2018cin, Staub:2019xhl, Heimel:2018mkt, Bothmann:2018trh, Luo:2017ncs, Chakraborty:2020yfc, Capozi:2019xsi}. The MPNN framework inherits the generality and powerfulness of Graph Neutral Network (GNN)~\cite{1555942,4700287}. It abstracts the commonalities between several of the most popular models for graph-structured data, such as spectral approaches~\cite{Kipf:2016gmz,2013arXiv1312.6203B,2016arXiv160609375D} and non-spectral approaches~\cite{2015arXiv150909292D} in graph convolution, gated graph neural networks~\cite{2015arXiv151105493L}, interaction networks~\cite{2016arXiv161200222B}, molecular graph convolutions~\cite{kearnes2016molecular}, deep tensor neural networks~\cite{2017NatCo...813890S} and so on~\cite{2018arXiv181208434Z}. In the MPNN, a collision event is represented as a numerical geometrical graph formed by a number of final state objects, which are non-linear models with a bunch of parameters that relates the output to the input graphs. The supervised learning is used to find optimized parameters, and will help to recognize the pattern in the collision events efficiently. Different from DNN, MPNN is a dynamic neural network and independent of the number and ordering of final state particles. Therefore, the MPNN is suitable for processing the graph representation of collision event. Recently, this method has been successfully applied to collider phenomenological studies, such as jet physics~\cite{Henrion2017NeuralMP}, Higgs physics~\cite{Ren:2019xhp} and supersymmetry~\cite{Abdughani:2018wrw}.

This paper is organized as follows. In Section~\ref{sec:event}, we describe the event generation and reconstruction for the signal and backgrounds. Next, in Section~\ref{sec:network}, we illustrate the event graph and network architecture for the MPNN approach. In Section~\ref{sec:results}, we present numerical results and discussions. Finally, we draw our conclusions in Section~\ref{sec:conclusion}.

\section{Event generation and reconstruction}\label{sec:event}

The signal and background events at parton level are generated with \texttt{MadGraph5\_aMC@NLO v2.6.1} \cite{Alwall_2014} with the default parton distribution function (PDF) set NNPDF2.3QED \cite{Ball:2013hta} at the LHC with leading order with center-of-mass energy $\sqrt{s}$ = 14 TeV. We employ the following cuts for parton level event generation : $p_{Tj} > 20$ GeV, $p_{Tb} > 20$ GeV, $p_{T\gamma} > 10$ GeV, $p_{T\ell} > 10$ GeV, $\eta_{j} < 5$, $\eta_b < 5$, $\eta_{\gamma} < 2.5$, $\eta_\ell < 2.5$, $\Delta R_{bb} < 1.8$, $\Delta R_{\ell\ell} < 1.3$, 70 GeV $< m_{jj}$ $<$ 160 GeV, 70 $<$ $m_{bb}< $ 160 GeV and $m_{\ell\ell} < $ 75 GeV, where $\ell$ denotes $e$ and $\mu{}{}$. We impose additionally  5 GeV $< m_{\ell\ell} < 75$ GeV for $jj\ell\ell\nu\bar\nu$, $\ell\ell b j$ and $tW+j$ backgrounds. The angular distance $\Delta R_{ij}$ is defined by
\begin{eqnarray}
\Delta R_{ij} = \sqrt{(\Delta\phi_{ij})^2+(\Delta \eta_{ij})^2} ~,
\label{eq:deltaR}
\end{eqnarray}
where $\Delta\phi_{ij} = \phi_i-\phi_j$ and $\Delta\eta_{ij}=\eta_i-\eta_j$ are the differences of the azimuthal angles and rapidities between particles $i$ and $j$, respectively.

The signal cross section is normalized to the next-to-next-to-leading-order (NNLO) accuracy in QCD \cite{Grigo:2014jma}, that is $\sigma_{gg \rightarrow hh}$ = 40.7 fb. The main background $t \bar t$ cross section is normalized to the NNLO QCD value 953.6 pb \cite{Czakon:2013goa}. Along with the signal and $t \bar t$, all other backgrounds and their normalized cross sections are listed in Table \ref{tab:xsec}.

We generate the low-$Q^2$ soft QCD pile-up events and apply hadronization via package \texttt{Pythia8243} \cite{Sjostrand:2014zea}, followed by detector simulation with \texttt{Delphes 3.4.2} \cite{deFavereau:2013fsa}. In the ATLAS card we consider the average amount of pile-up events per bunch-crossing as 100. We take the default parametrization implemented in the ATLAS card to distribute the hard scattering events and pile-up events randomly in time and $z$ positions. The maximum spread of pileup events in the beam direction is 0.25m and the maximum spread of pileup events in time is $8\times10^{-10}$ s.

In this work, we follow the default parametrization of \texttt{Delphes} ATLAS card to perform the pile-up subtraction and use the spatial vertex resolution parameter $|z|$ to perform charged pile-up subtraction. We consider every charged particle originating from a reconstructed vertex with $|z| > 0.01$ cm as coming from pile-up events and only keep those tracks that pass through the TrackPileUpSubtractor in \texttt{Delphes}.  

Similar to the tracks, the reconstructed jets are supposed to corrected from low-$Q^2$ pile-up events containing neutral particles. Jet pile-up subtraction is done via the JetPileUpSubtractor module that takes as input the jet constitutes and pile-up density $\rho$ based on the jet area. This technique helps to correct the jet momenta by calculating pile-up density ($\rho$) and jet area. Jets are clustered with the calorimeter tower elements using \texttt{Fastjet 3.3.2} \cite{Cacciari:2011ma} with anti-$k_T$ jet algorithm \cite{Cacciari:2008gp}, jet radius R = 0.4 with $p_T > 20 $GeV, and we allow the default estimation of $\rho$ with the calorimeter towers. As for the pile-up subtraction of missingET, we calculate it based on the pile-up subtracted jets, photons and leptons.

The Delphes card for ATLAS detector simulation is modified as :
\begin{itemize}
  \item Jets, including $b$-jets, with $p_T(j) > 20$ GeV and $|\eta_j| < 2.5 $ are selected.
  \item Flat $b$-tagging efficiency is $\epsilon_{b \rightarrow b} = 0.75$, mis-tagging efficiency for $c$ quark as $b$ is $\epsilon_{c \rightarrow b} = 0.1$, and mis-tagging rates of other jets are $\epsilon_{j \rightarrow b} = 0.01$ \cite{ATL-PHYS-PUB-2019-005}.
  \item Maximum transverse momenta ratio for lepton isolation is set as $\frac{\sum_{i \neq e} p_{Ti}}{p_{T \ell}} < 0.15$, where the sum is taken over the transverse momenta $p_{Ti}$ of all final state particles $i$, $i \neq \ell$, with $p_{Ti} > 0.5$ GeV and within angular distance  $\Delta R_{i\ell} < 0.3$ with lepton candidate $\ell$. Leptons with $p_T(\ell) > 10$ GeV and $|\eta_l| < 2.5 $ are selected.
  \item Isolation of photons also require  $\frac{\sum_{i \neq \gamma} p_{Ti}}{p_{T\gamma}} < 0.12$ for particles $i$, without including $\gamma$, with $p_{Ti} > 0.5$ GeV and within angular distance  $\Delta R_{i\gamma} < 0.3$ with photon candidate $\gamma$. Photons are required to have $p_T(\gamma) > 25$ GeV and $|\eta_\gamma| < 2.5 $ to be selected.
\end{itemize}

After the reconstruction, the missing transverse momentum $\mathbf{E}_{\rm T}^{\rm miss}$ is defined as the negative vector sum of the transverse momenta of the accepted photons, leptons and jets, and unused tracks as in \cite{Aaboud:2018tkc}:
\begin{equation}\label{eq:mpt}
  \mathbf{E}_{\rm T}^{\rm miss} = -  \sum_{\substack{\text{accepted} \\ \text{electrons}}} \mathbf{p}_{\rm T}^e  -   \sum_{\substack{\text{accepted} \\ \text{muons}}} \mathbf{p}_{\rm T}^\mu   -   \sum_{\substack{\text{accepted} \\ \text{photons}}} \mathbf{p}_{\rm T}^\gamma   -   \sum_{\substack{\text{accepted} \\ \text{jets}}} \mathbf{p}_{\rm T}^j    -   \sum_{\substack{\text{unused} \\ \text{tracks}}} \mathbf{p}_{\rm T}^{\rm track} ~ ,
\end{equation}
where the tracks with  $p_{T} > 0.4$ GeV and $|\eta| < 2.5 $ are considered.

We further apply the following cuts to reduce background events sufficiently relevant to the signals:
\begin{itemize}
  \item The two leading jets must be $b$-tagged, each with $p_T > 30$ GeV.
  \item Exactly two opposite sign leptons, each with $p_T > 25$ GeV.
  \item Modulus of  $\mathbf{E}_{\rm T}^{\rm miss}$ is required to be $E_{\rm T}^{\rm miss} > 20$ GeV.
  \item Angular distances for two leptons and for two $b$ jets are $\Delta R_{\ell \ell} < 1.0$ and $\Delta R_{b b} < 1.3$, respectively.
  \item Invariant masses for two leptons and for two $b$ jets respectively are $m_{\ell \ell} < 65 $ GeV and 95 GeV $ < m_{b b} < 140 $ GeV.
\end{itemize}
We export only the four momenta (also contain the corresponding charge signs of leptons and $b$-jet tagging information) of those events which passed the above cuts for later network training.

\section{Event graph and network architecture}\label{sec:network}

\begin{figure}
  \centering
  \includegraphics[width=8cm,trim=0 1.0cm 0 0,clip]{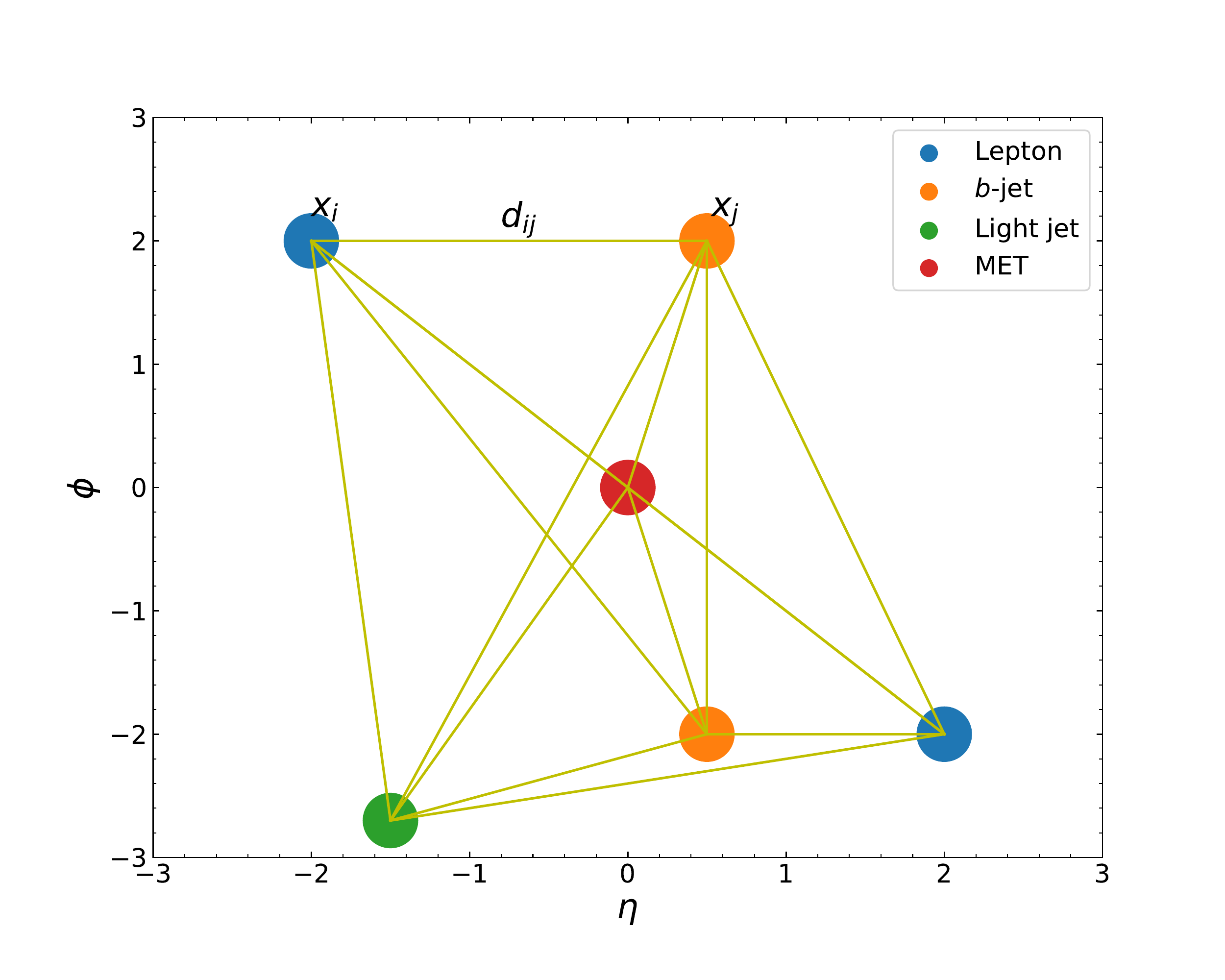}
  \includegraphics[width=8cm]{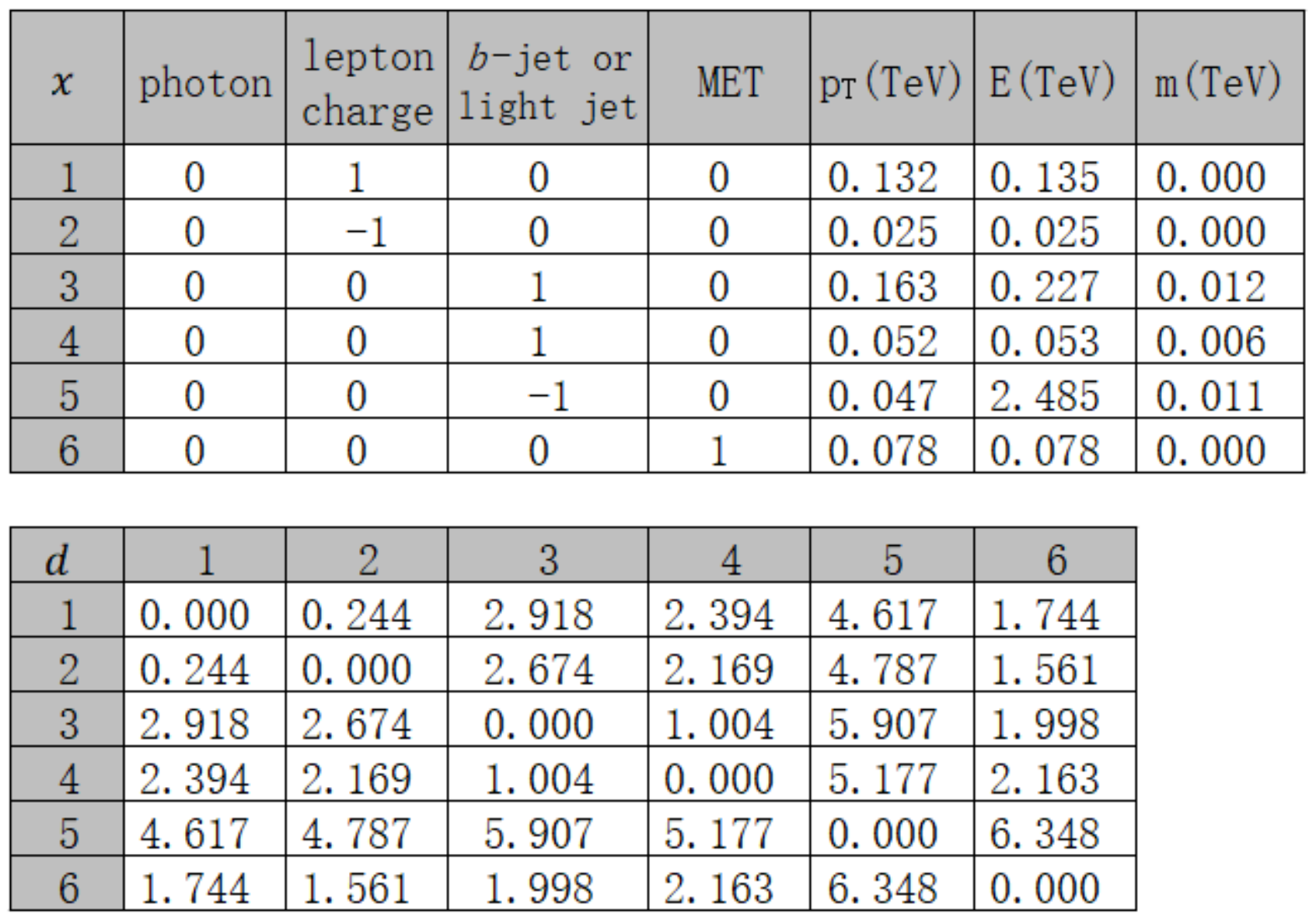}
  \caption{ The left figure illustrates an event graph, which includes nodes (circles) and edges (yellow lines), for a simulated signal event. A node represents a final state object passed all cuts and an edge represents the angular distance between two nodes. The upper right table shows the six objects; each of them is a seven-dimensional feature vector $x_i = (I_1, I_2, I_3, I_4, p_T, E, m)$ with $I_i$ features identifying its type, e.g., $I_1 = 1$ for a photon, $I_2$ is the charge of the corresponding lepton, $I_3 = 1$ is a $b$-tagged jet, $I_3 = -1$ is a non-$b$-tagged light jet, $I_4 = 1$ is the missing transverse momentum. $p_T$, $E$ and $m$ are the transverse momentum, energy and mass of the object. The table at the bottom shows the angular distances Eq. (\ref{eq:deltaR}) between a pair of nodes for all six objects.}\label{fig:graph}
\end{figure}

Each collider event obtained in the preceding section is converted to an event graph as the input for our neural network. Fig. \ref{fig:graph} illustrates a simulated signal event as an event graph which consists of nodes and edges. A node represents a final state object passed all the cuts and this object can be a photon, lepton, jet or missing transverse momentum (MET). Each node has a seven-dimensional feature vector $\bm x_i = (I_1, I_2, I_3, I_4, p_T, E, m)$ which contains the major property of the corresponding final state. For the elements of a feature vector, $p_T$, $E$ and $m$ are respectively the transverse momentum, energy and mass of the object, while the default values for $I_i$ are $0$, with $I_1 = 1$ for a photon, $I_2$ being the charge of the lepton, $I_3 = 1$ for a $b$-tagged jet, $I_3 = -1$ for a non-$b$-tagged jet, $I_4 = 1$ for the MET. Each pair of nodes are linked by an edge which is weighted by the angular distances (\ref{eq:deltaR}) between the corresponding two nodes.

Due to the rotation invariance of the differential cross section of the collider events around the beam axis, we can get rid of the information of azimuthal angle dependence of the event from the node features, and the difference of azimuthal angles is encoded in edge weights. This will make sure that the classification is not dependent on the definite azimuthal angle of the final states of an event, and stable w.r.t. the rotation of the event around the beam axis. The other two advantages of such an event graph design are: (1) The number of nodes equal to the number of final state objects, i.e., number of nodes is not fixed, which guarantees to use full information of final state objects; \footnote{We verified the assumption by restricting the number of light jets at the final states , and obtained best result by using full information.} (2) The node features and edge weights are easily transformed by the four momenta of the object, no sophisticated discriminants are needed to be constructed, which makes the model quite general and easy to implement to other scenarios as well.

\begin{figure}
  \centering
  \includegraphics[width=16cm,trim=1cm 0 0 0,clip]{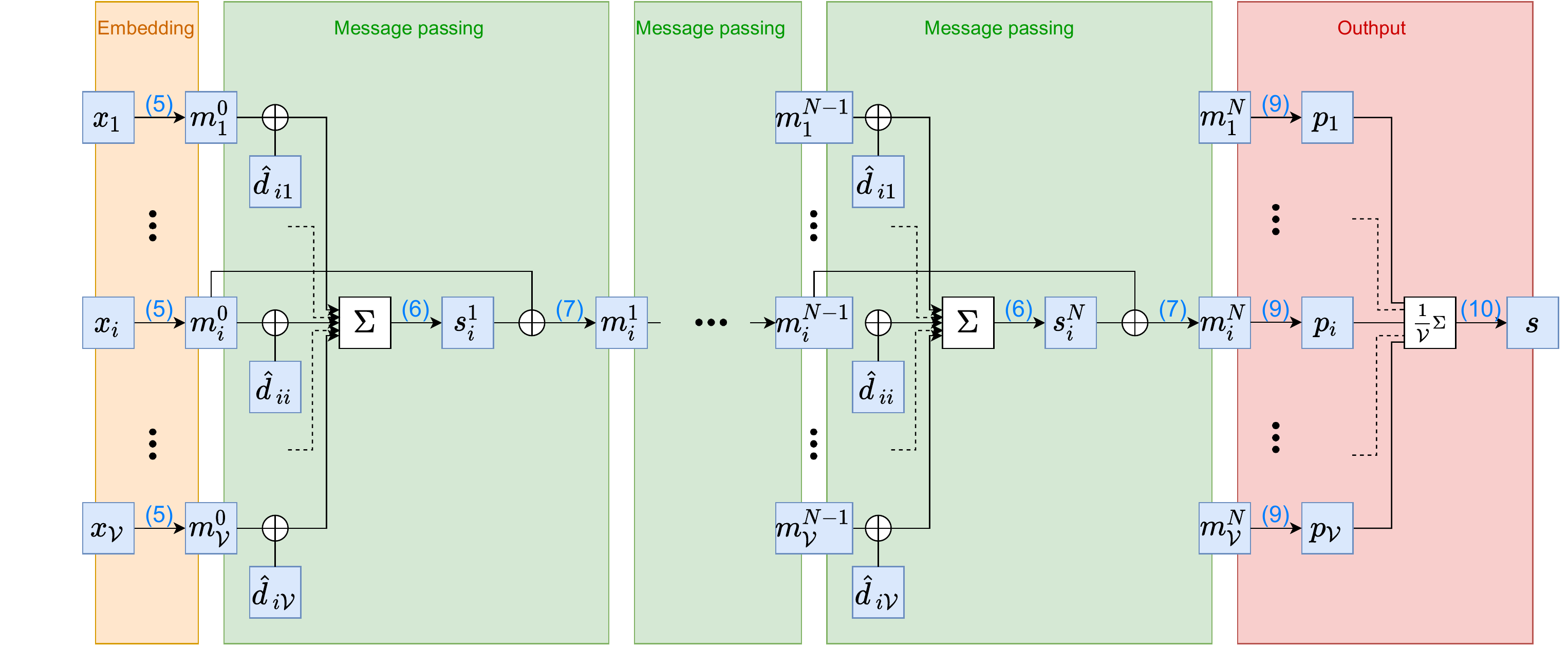}
  \caption{The schematic diagram of the MPNN classifier which consists of one embedding layer, $N$ pairs of message passing layers and one output layer (these layers are sequentially 
connected by non-linear transformations). The numbers in parentheses correspond to equation numbers in the text, the operators $\textcircled{+}$, $\Sigma$ and $\frac{1}{\mathcal{V}}$ denote vector concatenation, summation and averaging, respectively, the summation and averaging run over all $\mathcal{V}$ nodes. }\label{fig:network}
\end{figure}

The structure of our MPNN is shown in Fig. \ref{fig:network}, which consists of one embedding layer, $N$ message passing layers and one output layer. The embedding transformation for input data is given by,
\begin{equation}
\bm m^0_i = {\rm ReLU}(W_m^0 \bm x_i + \bm b_m^0) ~, \label{eq:embeding}
\end{equation}
where $W_m^0$ and $\bm b_m^0$ are learnable weight and bias vectors, 
and the activation function ReLU is the rectified linear unit \cite{nair2010rectified}. 
The dimension of $\bm m^0_i$ is higher than $\bm x_i$. It can be seen that $i$-th node $m^0_i$ in the embedding layer is a vector which only contains information from input feature $x_i$ without including any geometrical pattern of event graph. Then, the $i$-th node in the $n$-th message passing layer is obtained by the following transformation,
 \begin{eqnarray}
	&& \bm{s}_i^n = \sum_j \mathrm{ReLU}\left(W_m^n\, (\bm{m}_j^{n-1}\textcircled{+}\hat{\bm{d}}_{ij}) + \bm{b}_m^n\right), \label{eq:mp} \\ 
	&& \bm{m}_i^n = \mathrm{ReLU}\left(W_s^n\, (\bm{m}_i^{t-1}\textcircled{+}\bm{s}_i^n) + \bm{b}_s^n\right) ~, \label{eq:update}
\end{eqnarray}
where $i$ and $j$ are indices of nodes, $\bm s_i^n$ is intermediate vector, $\textcircled{+}$ represents vector concatenation, $W$s and $\bm b$s are learnable weights and biases. The message passing process is realized by two sub-processes: first, Eq. \ref{eq:mp}  
collects information from all previous nodes and distances between nodes; 
second, Eq. \ref{eq:update} passes this information together with previous node 
to the next one. By repeating this process, each note in the message passing layer 
gets knowledge of other nodes and relationships between them and updates itself. Therefore, the message-passing mechanism is the key for automatically extracting features of the input event graph, which efficiently disseminates the information among all the nodes taking into account the connections between nodes. After $N$ iterations, each node state vector can be viewed as an encoding of the whole event graph representing the whole information of both the kinematic features of all final states and the geometrical relationship between them. Here, we expand edge weight $d_{ij}$ onto 21 Gaussian bases to make it more suitable for linear transformation \cite{Abdughani:2018wrw}, and the $k$-th component of this weight vector is
\begin{equation}
	(\hat{\bm{d}}_{ij})_k = \exp \left\{ \frac{(d_{ij} - \mu_k)^2}{2 \sigma^2} \right\} ~,
\end{equation}
where $\mu_k$ is linearly distributed in range of [0, 5] and $\sigma$ = 0.25. Such an expansion is inspired by radial basis function networks \cite{broomhead1988radial,SCHWENKER2001439} which can solve non-linear problems by mapping input into high dimensions.

At the output layer, we use the sigmoid function on the vector $m_i^N$ to get the probability $p_i$ of the node $i$ as
\begin{equation}
	p_i = \sigma(W_p \bm m_i^N + b_p) = \frac{1}{1 + e^{-(W_p \bm m^N_i +  b_p)}} ~. \label{eq:sigmod}
\end{equation}
and then average the probabilities from all nodes at the output layer by
\begin{equation}\label{prob}
	s = \frac{1}{\mathcal{V}} \sum p_i ~,
\end{equation}
with $\mathcal{V}$ being the number of nodes in the input event which is the number of 
final state particles in an event. It should be mentioned that $\mathcal{V}$ is not a 
constant, e. g., if there are two extra light jets and one photon in an event apart 
from the required two $b$-jets, two leptons and one MET, then we have $\mathcal{V}$ = 7.

The MPNN can be efficiently trained using supervised learning method. We adopt binary-cross-entropy as the loss function. Although increasing the number of hidden layers can 
enable the network to lean more complex features in the data, it may have disadvantages 
like overfitting and time-consuming. We find that for our network $N$=3 is the most optimal choice \footnote{Message passing layer with $N=3$ can increase significance by about 5\% 
compared to $N=2$ , while $N=4$ can only increase significance by less than 1\% 
compared to $N=3$}. $W_m^0$, $W_m^n$, $W_s^n$, and $W_p$s in Eq. (\ref{eq:embeding} - \ref{eq:update},~\ref{eq:sigmod}) are 30$\times$7, 30$\times$51, 30$\times$60 and 1$\times$30 matrices, respectively. Thus, the overall number of learnable parameters in our MPNN model is 10441. The Adam \cite{Kingma:2014vow} optimizer with a learning rate of 0.001 is used to optimize the model parameters based on the gradients calculated on mini-batch of 128 training examples. A separate set of validation examples is used to measure the generalization performance while training to prevent over-fitting using the early-stopping technique. All these are implemented in the deep learning framework of PyTorch \cite{NIPS2019_9015} with CUDA platform and trained on a NVIDIA Titan Xp GPU with 12 Gb DDR5 memory for acceleration. One cycle of training and validation takes about half an hour when the size of the training data set and the validation data set are 300k and 100k, respectively. Note that signal and backgrounds have equal training and testing samples, while each sub-background has a number of samples proportional to cross section after the baseline cuts, e.g. 1.8568/2.2178 $\times$ 150K training samples for $t\bar t$, 0.2189/2.2178 $\times$ 150K for $tW+j$, and so on, where sum of the cross sections of all backgrounds after the baseline cuts is 2.2178 fb (see Table. \ref{tab:xsec}).

\section{Results and discussions}\label{sec:results}

In order to estimate the observability of the signal, we calculate the signal significance ($\alpha$) with the following formula,
\begin{align}
\label{alpha}
\alpha=S/\sqrt{B+(\beta B)^2},
\end{align}
where $S$ and $B$ denote number of signal and background events after our selections, respectively. ${\cal L}$ is the integrated luminosity of the collider. It should be mentioned that the main systematic uncertainty is parameterized by the factor of $\beta$ in our calculations.

\begin{figure}
\centering
\includegraphics[width=1.0\textwidth]{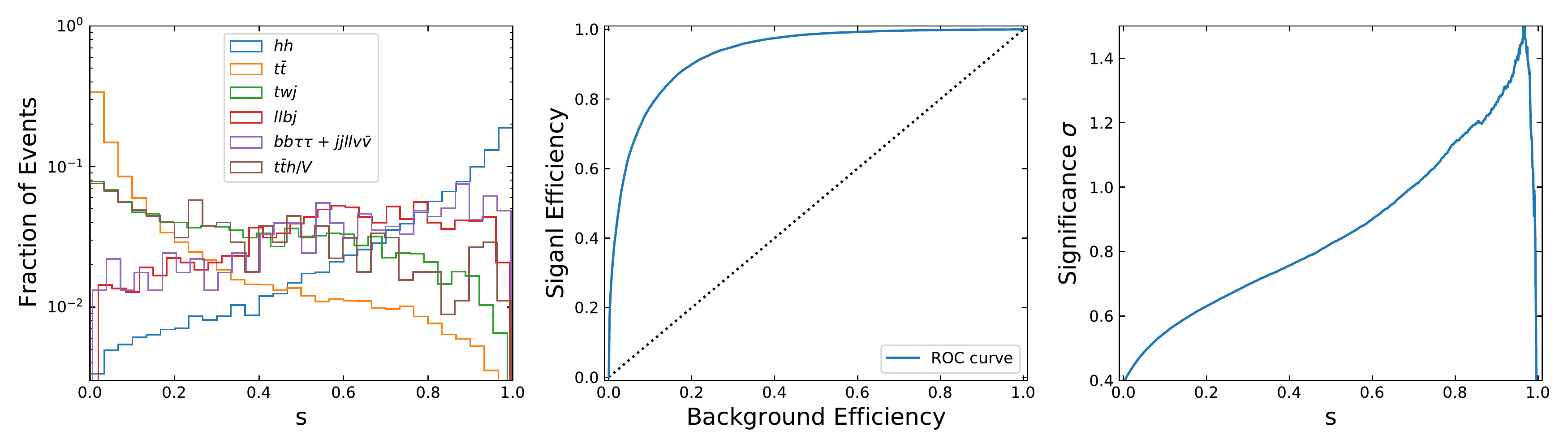}
\caption{The MPNN training results for the signal ($hh$) and backgrounds ($t\bar{t}$, $tWj$, $\ell^+\ell^- bj$, $b\bar{b}\tau\tau+jj\ell^+\ell^- \nu\bar{\nu}$ and $t\bar{t}h/V$) in the SM at 14 TeV LHC. {\it Left panel:} the event fractions of signal and each background versus the final score $s$. {\it Middle panel:} the ROC curve of signal and background. {\it Right panel:} the signal significance versus the final score $s$. The luminosity ${\cal L}=3000$ fb$^{-1}$ is assumed.}\label{fig:MPNN}
\end{figure}

Firstly, we focus on the SM Higgs pair production process $pp \to hh \to b\bar{b}WW^* \to b\bar{b}\ell^+\ell^-+E^{miss}_T$ at 14 TeV LHC with the luminosity of 3000 fb$^{-1}$. In Fig.~\ref{fig:MPNN}, we show the output of the trained MPNN evaluated on the validation test. The left panel is the discrimination score $s$, i.e., the probability distribution in Eq. (\ref{prob}), for the signal and the background processes. We label the signal as ``1'' and the background as ``0'' before training. As expected, the signal peaks near the $s = 1$ and dominant background $t\bar{t}$ peaks near the score $s = 0$, which are well separated from each other. For a given value of score, $s_0$, we can add the signal or background events in the range of $[s_0, 1]$ in the left panel and then obtain the receiver operating characteristic (ROC) curve in the middle panel, where the signal and background efficiencies are the fraction of the survival events in the initial signal and background events, respectively. We can see that the ROC curve increase steeply and show a good discrimination in the signal and background. The right panel shows the significance of signal as a function of score. Unfortunately, the maximum value of the significance for the SM Higgs pair process $pp \to hh \to b\bar{b}WW^* \to b\bar{b}\ell^+\ell^-+E^{miss}_T$ can only reach about $1.5\sigma$ at the HL-LHC.

\begin{table}[ht]
  \caption{Signal and background cross sections in fb unit before hadron-level cuts, 
but after baseline cuts and after MPNN validation process requiring the signal events 
number $N_{\rm sig} = 20$ to have reasonable statistics. The significance $\alpha$ is calculated by using the Eq.~\ref{alpha} with $\beta=0$ for simplicity. }\label{tab:xsec}
\centering
\resizebox{\textwidth}{15mm}{
\begin{tabular}{|c|c|c|c|c|c|c|c|c|c|c|}
  \hline
   & $hh$ & $t\bar t$ & $tW+j$ & $\ell^+ \ell^- bj$ & $t \bar t h$ & $\tau^+ \tau^- b \bar b$ & $t \bar t V$ & $jj\ell^+ \ell^- \nu \bar{\nu}$ & $\alpha(\sigma)$ & $S/B$ \\
   \hline
  No cut & 40.7 \cite{Grigo:2014jma} & 953600 \cite{Czakon:2013goa} & 123200 & 117100 \cite{deFlorian:2018wcj} & 661.3 \cite{Dittmaier:2011ti} & 29070 \cite{deFlorian:2018wcj}& 1710 \cite{deFlorian:2016spz} & 48200 \footnote{Applied an NLO k-factor of 2.0.} & $\simeq 0$ & $\simeq 0$ \\
  Baseline cuts & 0.0105 & 1.8568 & 0.2189 & 0.0675 & 0.0247 & 0.0246 & 0.0153 & 0.0101 & 0.3876 & 0.0047 \\
  MPNN & 0.0067 & 0.0581 & 0.0180 & 0.0152 & 0.0080 & 0.0025 & 0.0018 & 0.0017 & 1.13 & 0.06 \\
  \hline
\end{tabular}}
\normalsize
\end{table}

At the last row of the Table~\ref{tab:xsec}, we give the sensitivity of the SM signal process $pp \to hh \to b\bar{b}WW^* \to b\bar{b}\ell^+\ell^-+E^{miss}_T$ for MPNN, at 14 TeV LHC with the luminosity of 3000 fb$^{-1}$. In order to guarantee the statistic, we require to have 20 signal events after all selections for each method. the signal significance given by MPNN is about 1.12 $\sigma$.

\begin{figure}
\centering
\includegraphics[width=8.1cm,height=9cm,trim=0 0 0 0,clip]{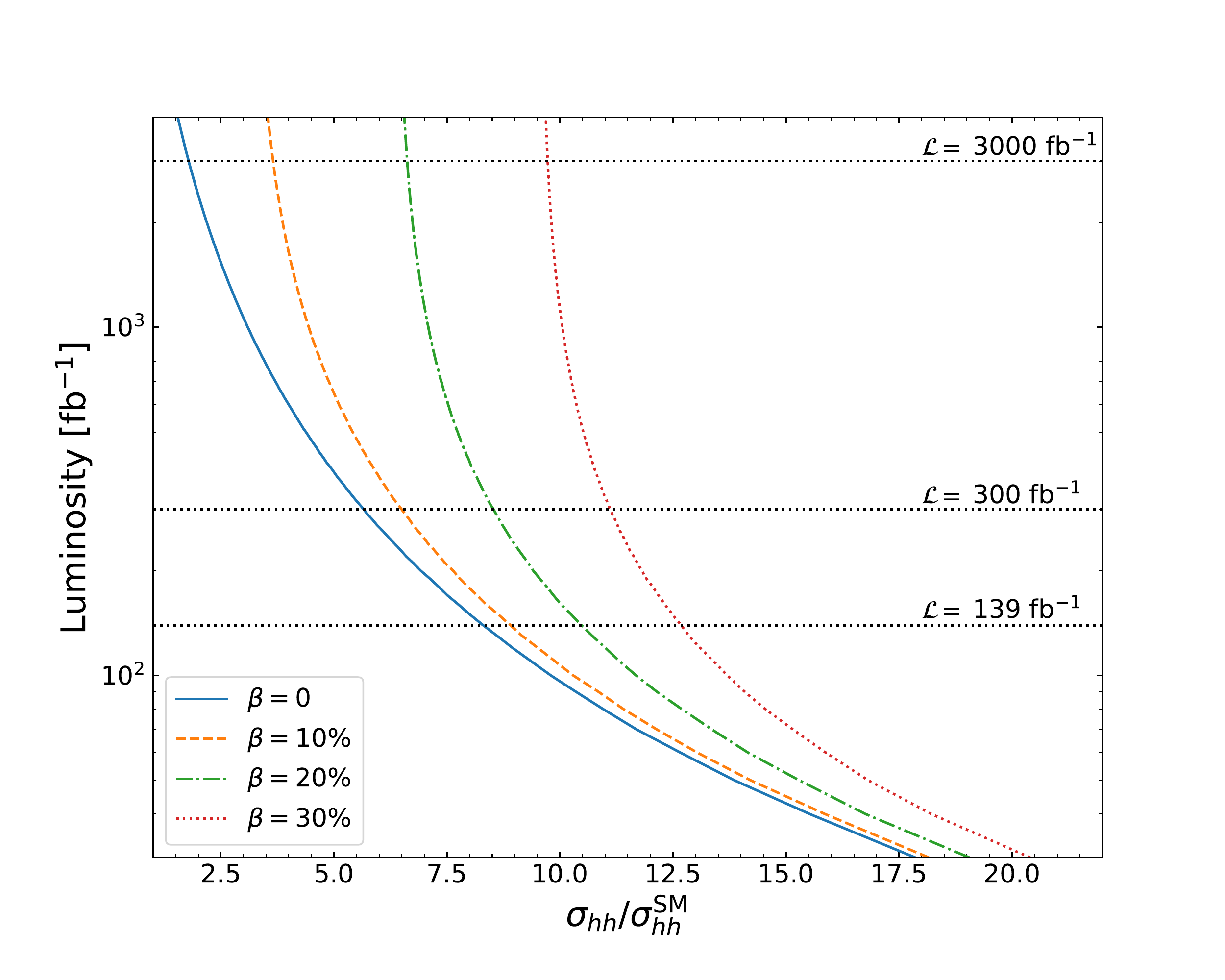}
\includegraphics[width=8.1cm,height=9cm,trim=0 0 0 0,clip]{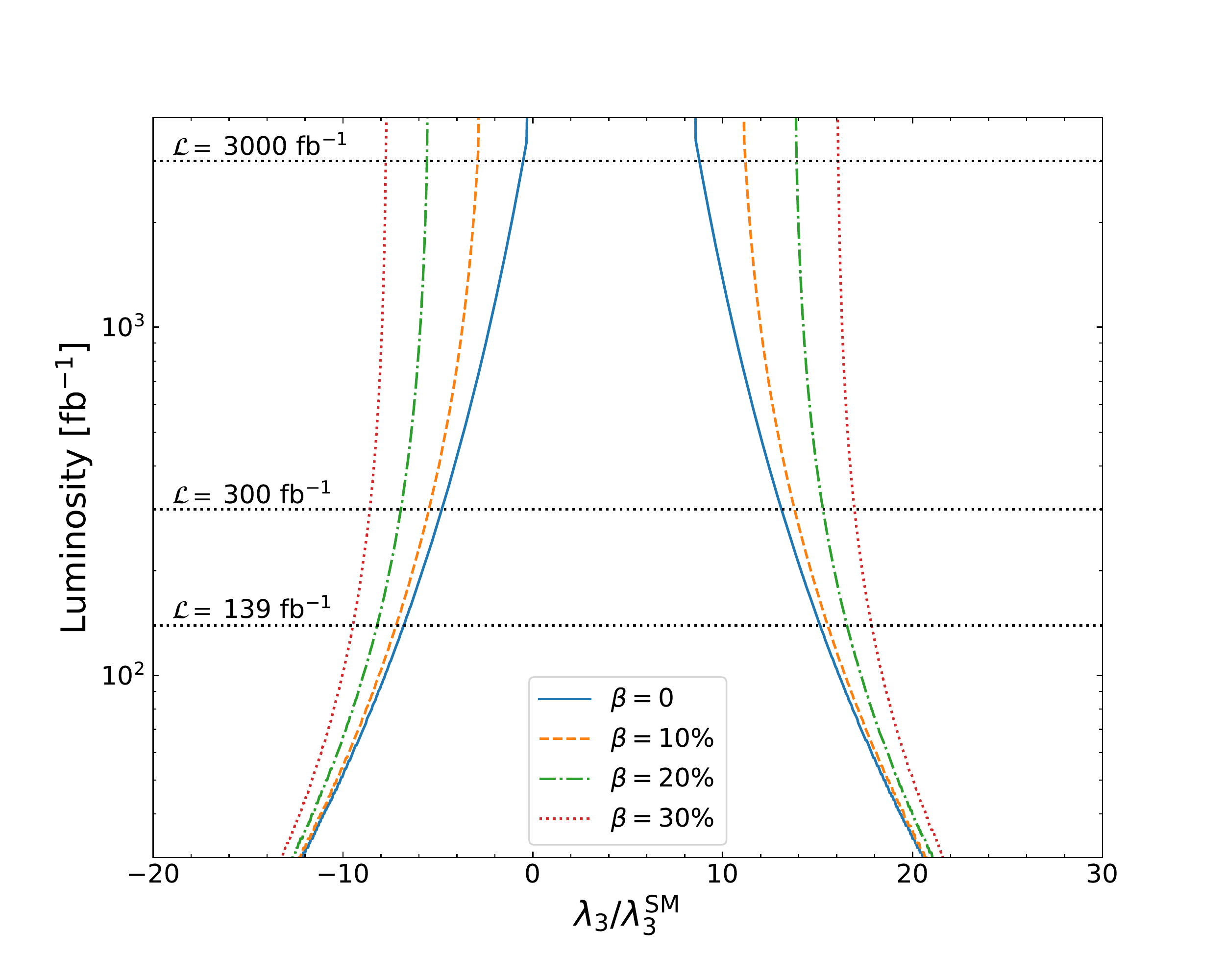}
\caption{ The $2\sigma$ upper bounds on production cross section of the Higgs pair ({\it left panel}) and triple Higgs coupling ({\it right panel}) at 14 TeV LHC.}\label{fig:k3}
\end{figure}

Finally, we apply our method to constrain the production cross section of the Higgs pair and the Higgs trilinear coupling in the BSM at 14 TeV LHC. We adopt the model-independent way to present the $2\sigma$ limits on the ratio of $\sigma_{hh} / \sigma^{SM}_{hh}$ in the left panel of Fig.~\ref{fig:k3}, where we take the systematic uncertainty $\beta=0, 10\%, 20\%, 30\%$ for example. It can bee seen that the production cross section of the Higgs pair larger than 13.5 times of the SM prediction can be excluded for the luminosity ${\cal L}=139$ fb$^{-1}$ and systematic error $\beta=30\%$. If $\beta$ can be controlled at 10\%, the $2\sigma$ upper bound on the ratio of $\sigma_{hh} / \sigma^{SM}_{hh}$ will be reduced to 9.5. Such results can be improved to be 10.2 for $\beta=30\%$ and 3.7 for $\beta=10\%$ at the HL-LHC. Provided $\beta=0$, this limit on $\sigma_{hh}/\sigma^{SM}_{hh}$ will become 1.5. Besides, we reinterpret these bounds for triple Higgs coupling in the right panel of Fig.~\ref{fig:k3}. We find that the ratio of $\lambda_{3h}/\lambda^{SM}_{3h}$ can be constrained to the range of $[-10,18]$ for ${\cal L}=139$ fb$^{-1}$ and $\beta=30\%$, and will be further narrowed down to the range of $[-3,11.5]$ for ${\cal L}=3000$ fb$^{-1}$ and $\beta=10\%$ at $2\sigma$ level.

\section{Conclusions}\label{sec:conclusion}
In this paper, we explored the discovery potential of Higgs pair production process $pp \to hh \to b{b}WW^* \to 2b + 2\ell + E_{\rm T}^{\rm miss}$ with the Message Passing Neural Network at the (HL-)LHC. In the MPNN, we can represent each collision event as an event graph that consists of the final state objects, and use the supervised learning to optimize training parameters. By using the MPNN, we obtained that the significance of the SM Higgs pair production process can reach the maximum of about $1.5\sigma$ at the HL-LHC. Then, we extended our study to constrain the production cross section of the non-resonant Higgs pair and the triple Higgs trilinear coupling in a model-independent way. We found that the production cross section of the Higgs pair larger than 10.2 times of the SM prediction can be excluded at $2\sigma$ level for the HL-LHC when a  30\% systematic uncertainty is included. If the systematic error can be well controlled, such as 10\%, this upper bound can be improved to 3.7 times of the predicted by the SM, which will constrain the triple Higgs coupling to the range of $[-3,11.5]$. Therefore, we expect this channel can play an important role in enhancing the sensitivity of the combining analysis of SM Higgs pair production at the HL-LHC .

\section*{Acknowledgments}

This work was supported by the National Natural Science Foundation of China (NNSFC) under grant
Nos. 12047560, 11705093, 12075300, and 11851303.

\bibliography{refs}

\end{document}